\documentclass[%
 aps,
 prb,
 amsmath,amssymb,
 showpacs,
 reprint,%
]{revtex4-1}

\usepackage{graphicx}
\usepackage{dcolumn}
\usepackage{bm}
\usepackage{verbatim}

\begin{document}

\title{Trion formation and unconventional superconductivity in a three-dimensional model with short-range attraction}

\author{Pavel Kornilovitch}
 \email{pavel.kornilovich@gmail.com}
 \affiliation{Department of Physics, Oregon State University, Corvallis, Oregon 97331 USA} 

\date{\today}  

\begin{abstract}

A three-fermion problem in a three-dimensional lattice with anisotropic hopping is solved by discretizing the Schr\"odinger equation in momentum space. Interparticle interaction comprises on-site Hubbard repulsion and in-plane nearest-neighbor attraction. By comparing the energy of three-fermion bound clusters (trions) with the energy of one pair plus one free particle, a trion formation threshold is accurately determined, and the region of pair stability is mapped out. It is found that the ``close-packed'' density of fermion pairs is highest in a strongly anisotropic model. It is also argued that pair superconductivity with the highest critical temperature is always close to trion formation, which makes the system prone to phase separation and local charge ordering.   

\end{abstract}



\maketitle

{\em Introduction.}---  
Interest in fermion models with short-range attractive interactions has been growing since the discovery of high-temperature superconductors in 1986. Early work was summarized in the review.~\cite{Micnas1990} The authors argued that in complex solids such as multicomponent oxides there exist many bosonic degrees of freedom (phonons, magnons, polarization waves, and other types) that facilitate a non-retarded attraction between the carriers, which in some cases can overscreen the direct Coulomb repulsion at finite separation distances. It may lead to real-space pairing and potentially superconductivity if the pairs undergo Bose-Einstein condensation. Since the real systems are overly complex, it makes sense to introduce simpler phenomenological models that capture the essential physics. That leads to a family of $UV$ fermion models that comprise a short-range repulsion $U$ (typically of Hubbard type) and a finite range attraction $V$ (typically between nearest neighbors). In the field of unconventional superconductivity, these models occupy an intermediate place between fully microscopic models that include details of the original carrier-boson interactions~\cite{Micnas1990,Alexandrov1994,Scalapino2012} and the oversimplified phenomenological models of charged Bose gas.~\cite{Ogg1946,Schafroth1954,Schafroth1957} More recently, $UV$-type models were applied to trapped ions in optical lattices.~\cite{Jaksch2005,Bloch2008,Deng2011,Ohgoe2011,Ng2015}       

The main advantage of the $UV$ model relative to microscopic models is simplicity. Two-particle states in the $UV$ model~\cite{Alexandrov1993,Kornilovitch1995,daVeiga2002,Kornilovitch2004,Bak2007} and in the related dilute $t-J$ model~\cite{Emery1990,Lin1991,Petukhov1992,Kagan1994} were studied by many authors and in most cases the pairing threshold was derived analytically. In general, pair formation is now well understood. Much less is known about phase separation in those models. At a sufficiently large $V$, all particles in the system should form one big immobile cluster. (In the attractive fermion Hubbard model, clustering is prevented by the exclusion principle.) A physically relevant question is whether there exists a {\em finite} interval of intermediate $V$ strong enough to form pairs but not strong enough to cause phase separation. Then a liquid of real-space pairs could be stable.     

Answering the above question amounts to solving a many-fermion $UV$ problem which has not been done yet. Recently, progress was made by analyzing {\em three}-particle $UV$ clustering~\cite{Kornilovitch2013,Kornilovitch2014} and the following picture has emerged. The ground state of the {\em two}-particle system is a spin singlet with a nodeless coordinate wave function. When a third fermion attempts to bind to an existing pair the wave function must form a node. The node is equivalent to an effective repulsion $U_e$ that is larger than the dynamic repulsion $U$. Thus, attraction $V$ can overcome $U$ to form a stable pair and at the same time not being able to overcome $U_e$ to form a trion. As a result, the pair will repel the third fermion. In other words, the region of pair stability is protected by the fermion exclusion principle. The above picture was supported by exact solutions of the three-fermion Schr\"odinger equation in one~\cite{Kornilovitch2013} and two~\cite{Kornilovitch2014} spatial dimensions. A finite region of pair stability (in this case bipolarons) was also found in a $d = 1$ model with a long-range electron-phonon interaction by solving four- and six- fermion problems variationally.~\cite{Chakraborty2014} The present author is unaware of any other relevant work in this area.      

The goal of this paper is to extend the analysis of Ref.~[\onlinecite{Kornilovitch2014}] to $d = 3$. It turns out that the set of integral equations, to which the Schr\"odinger equation can be reduced by the procedure explained below, is the same in $d = 2$ and $d = 3$ as long as attraction $V$ is limited to {\em in-plane} nearest neighbors. The only difference in $d = 3$ is a three-dimensional rather than two-dimensional one-particle dispersion $\varepsilon_{\bf k}$ and a three-dimensional rather than two-dimensional Brillouin zone. Although the integral equations are harder to solve in $d = 3$ than in $d = 2$, the associated challenges are purely technical and have been overcome with more memory and compute time. The reward is new insights into the physics of real-space pairs. Similarly to $d = 1,2$ and consistent with the qualitative argument given above, the $d = 3$ $UV$ model is also found to possess a finite region of pair stability. Further, the pairs are ``most compact'' in a highly anisotropic version of model, which has implications for unconventional superconductivity. Based on those results, it is argued that optimal preformed-pair superconductivity is always close to phase separation.

\begin{figure}[t]
\includegraphics[width=0.46\textwidth]{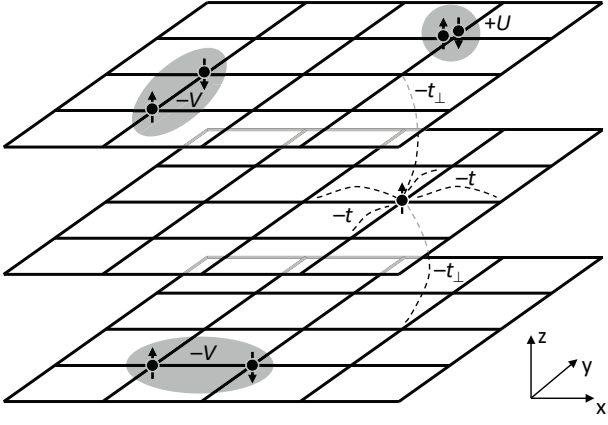}
\caption{Tetragonal $UV$ model with {\em in-plane} attraction.}
\label{3UV3d:fig:one}
\end{figure}

{\em Model.}---  
The {\em tetragonal} $UV$ model with in-plane attraction is defined by the Hamiltonian:
\begin{eqnarray}
H  & = & - t \sum_{{\bf m}, {\bf b}, \sigma} 
             c^{\dagger}_{{\bf m} \sigma} c_{{\bf m} + {\bf b}, \sigma}   
         - t_{\perp} \sum_{{\bf m}, {\bf b}_{\perp}, \sigma} 
             c^{\dagger}_{{\bf m} \sigma} c_{{\bf m} + {\bf b}_{\perp}, \sigma}   
\nonumber \\    
    &  & + \frac{U}{2} \sum_{\bf m} \hat{n}_{\bf m} \left( \hat{n}_{\bf m} - 1 \right) 
         - \frac{V}{2} \sum_{{\bf m}, {\bf b}} \hat{n}_{\bf m} \hat{n}_{{\bf m} + {\bf b}} \: .     
\label{3UV3d:eq:one}    
\end{eqnarray}
Here, $c^{\dagger}$ and $c$ are spin-$\frac{1}{2}$ fermion operators, {\bf m} numbers lattice sites, ${\bf b} = \pm {\bf x}, \pm {\bf y}$ numbers the four nearest neighbors within the $xy$ plane, ${\bf b}_{\perp} = \pm {\bf z}$ are the two nearest lattice neighbors across the planes, $\sigma = \pm \frac{1}{2}$ is the $z$-axis spin projection, and $\hat{n}_{\bf m} = \sum_{\sigma} c^{\dagger}_{{\bf m} \sigma} c_{{\bf m} \sigma}$ is the total fermion number operator on site ${\bf m}$. The kinetic energy is defined by in-plane and between-the-planes hopping amplitudes $t$ and $t_{\perp}$, see Fig.~\ref{3UV3d:fig:one}. The one-particle dispersion is
\begin{equation}
\varepsilon_{\bf k} = - 2t \left( \cos{k_x} + \cos{k_y} \right) - 2 t_{\perp} \cos{k_z} \: .  
\label{3UV3d:eq:two}
\end{equation}
Although Hamiltonian (\ref{3UV3d:eq:one}) is well defined for arbitrary $U$ and $V$, the present paper is focused on $U > 0, V > 0$.

\begin{figure}[t]
\includegraphics[width=0.48\textwidth]{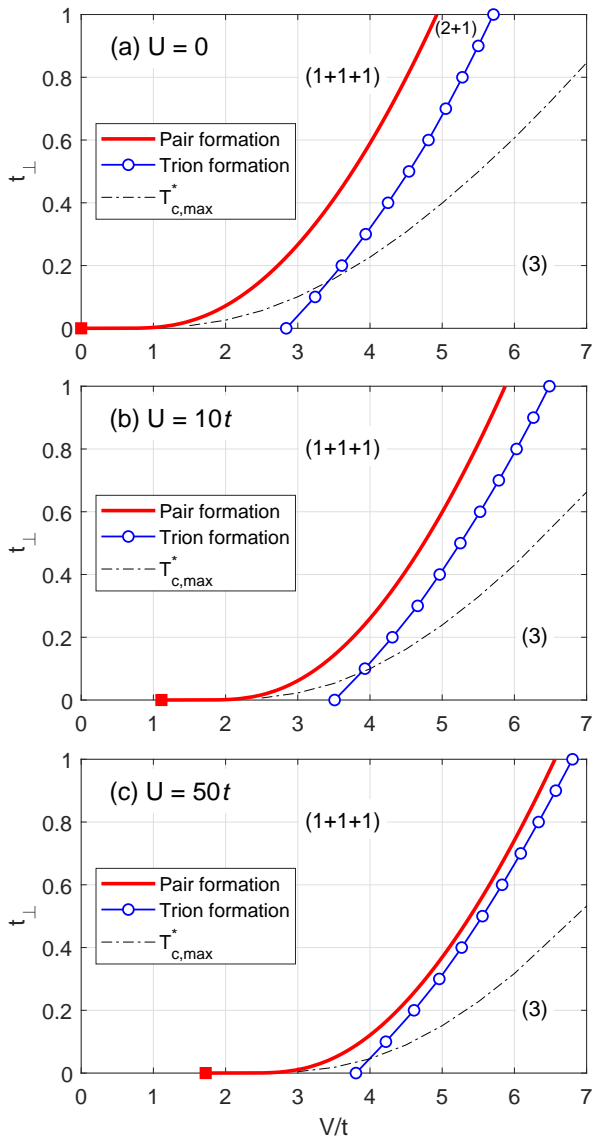}
\caption{(Color online.) Phase diagram of the tetragonal $UV$ model for ${\bf P} = 0$ and several $U$. The pair formation line is obtained from the exact two-fermion solution, Eq.~(\ref{3UV3d:eq:three}). The open circles are $S = 1/2$ trion formation threshold obtained from the three-fermion solution. The pair is stable between the pair formation and trion formation lines. Optimal pair superconductivity is found near the intersection of the $T^{\ast}_{c,{\rm max}}$ and trion formation lines. } 
\label{3UV3d:fig:three}
\end{figure}

The two-fermion case of model~(\ref{3UV3d:eq:one}) was investigated in Ref.~[\onlinecite{Kornilovitch2015}]. Singlet pair energy $E_2({\bf P})$, where ${\bf P}$ is pair momentum, follows from the exact dispersion equation, see Supplemental Material~\cite{SupplMat2019}
\begin{equation}
\left\vert \begin{array}{ccc}
M_{000} + \frac{1}{U}  & 2 M_{100}                        &  2 M_{010}                \\
M_{100}           &      M_{000} + M_{200} - \frac{1}{V}  &  2 M_{110}                \\
M_{010}           &      2 M_{110}                        &  M_{000} + M_{020} - \frac{1}{V}  
\end{array} \right\vert = 0 \: ,
\label{3UV3d:eq:three} 
\end{equation}
\begin{equation}
M_{nml} = \! \int^{\pi}_0 \!\!\!\! \int^{\pi}_0 \!\!\!\! \int^{\pi}_0  
\! \frac{dx dy dz}{\pi^3} \frac{\cos{(nx)} \cos{(my)} \cos{(lz)}}
{\vert E_2 \vert - a \cos{x} - b \cos{y} - c \cos{z}}  ,
\label{3UV3d:eq:four}
\end{equation}
where $a \equiv 4 t \cos{\frac{P_x}{2}}$, $b \equiv 4 t \cos{\frac{P_y}{2}}$, $c \equiv 4 t_{\perp} \cos{\frac{P_z}{2}}$. By setting $E_2 = - a - b - c$, the pair binding condition is obtained. For example, critical attraction strength $V_{\rm cr}$ can be expressed via $t_{\perp}$ for given $U$ and ${\bf P}$. This curve is shown in Fig.~\ref{3UV3d:fig:three} as the ``Pair formation'' line. Another important quantity supplied by the exact two-particle solution is pair effective radius $r^{\ast}$. It will be discussed below in relation to a maximum critical temperature attainable in a system of real-space pairs. 

The three-particle sector of Eq.~(\ref{3UV3d:eq:one}) is much harder to analyze. In 1986, Rudin,~\cite{Rudin1986} using methods developed earlier by Mattis,~\cite{Mattis1986} reduced the three-{\em boson} problem to a set of five integral equations (in $d = 2$) but did not proceed to solve them. More recently, the present author, using a similar integral equation method, solved the three-fermion case first in $d = 1$, Ref.~[\onlinecite{Kornilovitch2013}], and then in $d = 2$, Ref.~[\onlinecite{Kornilovitch2014}]. As mentioned earlier, the equations remain unchanged in $d = 3$ as long as attraction $V$ is still confined to planes. Numerical solution of the three-fermion tetragonal $UV$ model is presented below.

\begin{figure}[t]
\includegraphics[width=0.48\textwidth]{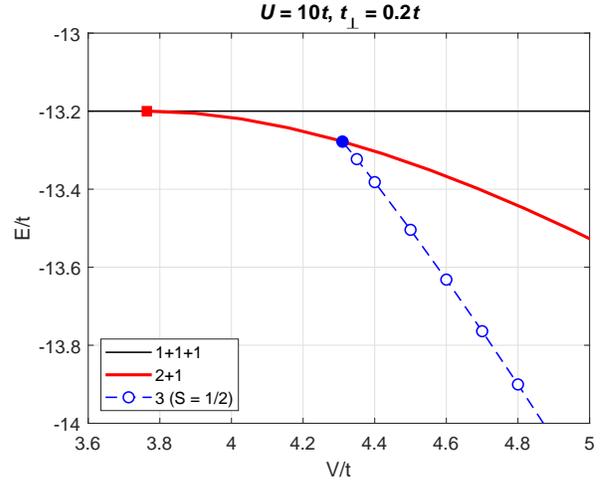}
\caption{(Color online.) Three-particle energies for $U = 10 \, t$, $t_{\perp} = 0.2 \, t$, and ${\bf P} = 0$. The thick solid line is the lowest energy of one singlet pair plus one free fermion. The solid square marks pair binding threshold, $V_{\rm cr} = 3.7636$. The open circles are computed energies of the lowest $S = 1/2$ trion state, which is doubly degenerate. The dashed line is guide to the eye. Extrapolation of the dashed line to the solid line yields a trion formation threshold of $4.31 t$, marked by a filled circle.}
\label{3UV3d:fig:two}
\end{figure}

{\em Method.}---  
The Schr\"odinger equation for the three-fermion wave function $\phi_{{\bf q}_1, {\bf q}_2, {\bf q}_3}$ is {\em nine}-dimensional, so direct solution is not practical. Fixing total momentum ${\bf P} = {\bf q}_1 + {\bf q}_2 + {\bf q}_3$ leaves only two three-dimensional variables ${\bf q}_1$ and ${\bf q}_2$. Next, since $V$ is of finite radius, the interaction part of the Schr\"odinger equation contains a finite number of integrals 
\begin{equation}
F_{i}({\bf q}) = \frac{1}{N} \sum_{\bf k} f({\bf k}) \phi_{{\bf k},{\bf q},{\bf P}-{\bf q}-{\bf k}} \: , 
\label{3UV3d:eq:six}
\end{equation}
with different permutations of $\phi$'s arguments and $f({\bf k}) = \cos{(\bf kb)}$, $\sin{(\bf kb)}$, or $1$, see Supplemental Material for details.~\cite{SupplMat2019} Expressing $\phi$ as a linear combination of $F_{i}$ and substituting back in Eq.~(\ref{3UV3d:eq:six}) results in a set of nine coupled integral equations for nine $F_{i}({\bf q})$. Thus, one two-variable function $\phi$ is replaced by nine one-variable functions $F_{i}$. The resulting equations occupy several pages and are not given here. In full form, they are written in Supplemental Material.~\cite{SupplMat2019} To solve the equations, the Brillouin zone is discretized into $12 \times 12 \times 12 = 1728$ points, ${\bf k}$ integrals are replaced by finite sums utilizing the three-dimensional Simpson integration rule, the entire set is transformed into a dense $(15,552 \times 15,552)$ matrix equation, and the system's energy $E$ is found via eigenvalue search. More details on this reduction methodology can be found in Refs.~[\onlinecite{Rudin1986,Kornilovitch2013,Kornilovitch2014,Mattis1986}]. 

At large $V$, three fermions always form a bound cluster, a trion. If $V$ is systematically reduced, eventually the trion energy becomes equal to the minimum energy of a bound pair plus one free fermion (with the same total momentum ${\bf P}$). By careful extrapolation of $E(V)$, trion formation threshold $V_3$ is determined. The procedure is illustrated in Fig.~\ref{3UV3d:fig:two}. In the $t_{\perp},V$ phase diagram, function $V_3(t_{\perp})$ defines a ``Trion formation'' boundary line that separates the region of stable pairs from the region of stable trions, see Fig.~\ref{3UV3d:fig:three}. By comparing the $12^3$ discretization with $12 \times 12 \times 8$ and $8^3$ discretizations, numerical errors in trion energies are estimated to be $< 0.05 \, t$, which is sufficient for determining the phase boundary with a $V$ uncertainty of $< 0.1 \, t$.

{\em Results.}---  
Trion physics in the $UV$ model is rich.~\cite{Kornilovitch2014} There are twelve states with total spin $S = 1/2$ and six more states with $S = 3/2$. At ${\bf P} = 0$, some states are double-degenerate but at ${\bf P} \neq 0$, they split into eighteen separate bands. Each trion state has its own binding threshold. In addition, there is a Nagaoka transition at large $U$ and $V$.~\cite{Kornilovitch2014} The present paper is concerned with only one aspect of this physics: formation of the lowest $S = 1/2$ trion at ${\bf P} = 0$, and the region of stability of {\em pairs} against three-fermion clustering. 

Figure~\ref{3UV3d:fig:three} shows the model's phase diagram at several $U$. The trion formation line is the main computational result of the paper. The line runs approximately parallel to the pair binding line except at very small $t_{\perp}$ where the pair line has a logarithmic singularity and separation is larger. Between the two lines pairs are stable. A finite region of stability is consistent with the exclusion-based repulsion mechanism described in Introduction. As $V$ is increasing, once a first nodeless state (singlet pair) is formed, it takes an additional {\em finite} increase of $V$ to form the next state with a node (trion). Being qualitative, the ar\-gu\-ment should be valid in a wide range of model parameters, which is supported by the observation that the phase diagrams at small and large $U$ are qualitatively similar. These are welcome results for real-space pairing. In mapping the complex interactions of real solids on pseudo-potentials $U$ and $V$, there is a finite chance that $U$ and $V$ may land in the region of pair stability without hyperfine tuning of parameters. One can conjecture that the pair stability region should also exist in four-fermion and many-fermion systems, again because the exclusion-based repulsion should be in effect there, too. The exact phase boundaries will be different, however. Determining them would require a rigorous solution of the many-fermion $UV$ case, which has not been done yet.    

Optimal superconductivity is discussed next. It has been argued~\cite{Kornilovitch2015} that the highest critical temperature $T^{\ast}_c$ of a system of real-space pairs corresponds to ``close packing'' of pairs. $T^{\ast}_c$ can be found by equating the pair density (expressed as a Bose integral over pair dispersion) to an inverse pair volume:  
\begin{eqnarray}
& & \frac{A}{\sqrt{[ 1 + ( r^{\ast}_x )^2 ] [ 1 + ( r^{\ast}_y )^2 ] [ 1 + ( r^{\ast}_z )^2 ] }} =
\nonumber \\
& & \hspace{1.0cm} = \int_{\rm BZ} \frac{d^3 {\bf P}}{(2\pi)^3} 
\frac{1}{\exp{\left\{ \frac{E_2({\bf P}) - E_0}{T^{\ast}_c} \right\} } - 1} \: , 
\label{3UV3d:eq:five}
\end{eqnarray}
where $A \approx 0.1$. It turns out that for given $V$ and $U$, $T^{\ast}_c$ as a function of $t_{\perp}$ has a maximum. Very small $t_{\perp} \rightarrow 0$ destroy three-dimensional quantum coherence driving $T^{\ast}_c \rightarrow 0$. $T^{\ast}_c$ must decrease with $t_{\perp}$ to keep the integral in Eq.~(\ref{3UV3d:eq:five}) finite. On the other hand, a large kinetic energy $t_{\perp} \rightarrow t$ unbinds pairs and causes $r^{\ast} \rightarrow \infty$. That implies a zero packing density and again $T^{\ast}_c \rightarrow 0$. Optimal $t_{\perp}$ as a function of $V$ is shown in Fig.~\ref{3UV3d:fig:three} as the $T^{\ast}_{c, {\rm max}}$ line. It turns out that $T^{\ast}_{c}$ also increases with $V$ along the $T^{\ast}_{c, {\rm max}}$ line.~\cite{Kornilovitch2015} Thus, an ``absolute maximum'' of $T^{\ast}_{c}$ is achieved at large $V$ and intermediate $t_{\perp}$ that is well beyond trion formation and deep in the phase separation regime. To avoid clustering, optimal pair superconductivity must be sought {\em inside} the pair stability region near the intersection of $T^{\ast}_{c, {\rm max}}$ and trion formation lines.

\begin{figure}[t]
\includegraphics[width=0.48\textwidth]{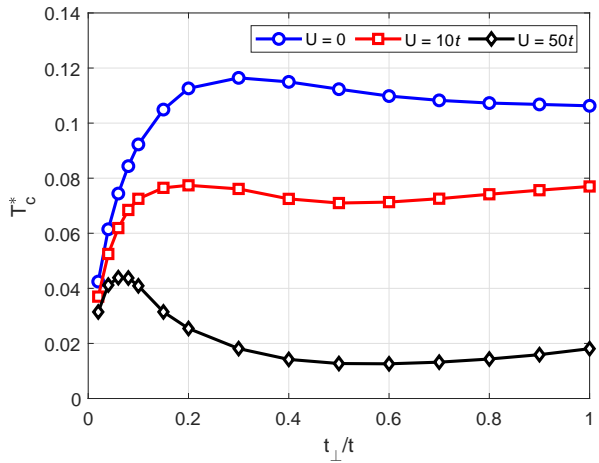}
\caption{(Color online.) The ``close packing'' critical temperature $T^{\ast}_c$ of Eq.~(\ref{3UV3d:eq:five}), computed along the trion formation line shown in Fig.~\ref{3UV3d:fig:three}. Note a maximum of the $U = 50t$ line near $t_{\perp} \approx 0.1 \, t$. }
\label{3UV3d:fig:four}
\end{figure}

Figure~\ref{3UV3d:fig:four} shows $T^{\ast}_c$ along the trion formation lines, computed using Eq.~(\ref{3UV3d:eq:five}). Note a broad maximum at $t_{\perp} \approx ( 0.1 - 0.3 ) t$. At large $U$, the maximum shifts to lower $t_{\perp}$ and becomes more pronounced. One concludes that in the presence of strong correlations the maximal achievable critical temperature is highest in anisotropic systems. Of course, pair superconductivity may exist in isotropic systems too, but systems with $\sim 10{\rm x}$ anisotropy offer the best balance between compact pairs (high packing density) and small pair masses. Perhaps, this is why anisotropic cuprates and pnictides have higher $T_c$ than Ba$_{0.6}$K$_{0.4}$BiO$_3$ and other isotropic oxides.   

Another important conclusion that can be drawn from the above analysis is that optimal pair superconductivity is {\em always} close to clustering and, more generally, to phase separation. There is ample evidence of charge ordering in the cuprates and pnictides in the form of charge density waves,~\cite{Chang2012,Ghiringhelli2012,Comin2015} stripes,~\cite{Tranquada1995,Kivelson2003,Foerst2014,Li2016} and nematic order.~\cite{Chu2010,Yi2014,Kim2014} This effect finds a natural explanation in the $UV$ model if charge ordering is regarded as a form of phase separation. As shown above, to maximize $T_c$ the system should be driven to the brink of clustering but without crossing the threshold. However, local spatial fluctuations can throw the system over the threshold locally, thus creating a variety of local clusters that experimentally manifest themselves as charge order. This subject deserves deeper investigation.

\begin{acknowledgments}

The author wishes to thank James Hague and Ganiyu Adebanjo for helpful discussions on the topics of this paper.     

\end{acknowledgments}


\begin{widetext}

\section{\label{3UV3d:sec:app}
Supplemental Material
}

Working equations for the two-fermion and three-fermion cases of the tetragonal $UV$ model are derived below.

\subsection{\label{3UV3d:sec:app:a}
Two fermions with total spin $S = 0$.
}

Schr\"odinger equation for a {\em symmetrized} wave function $\phi_{{\bf k}_1{\bf k}_2} = \phi_{{\bf k}_2{\bf k}_1}$ reads
\begin{equation}
( E_2 - \varepsilon_{{\bf k}_1} - \varepsilon_{{\bf k}_2} ) \, \phi_{{\bf k}_1 {\bf k}_2}
= U \frac{1}{N} \sum_{\bf q}  \phi_{{\bf q}, {\bf k}_1 + {\bf k}_2 - {\bf q} }  
- V \sum_{{\bf b}_{+}} \frac{1}{N} \sum_{\bf q}     
\left[ e^{ i ( {\bf q} - {\bf k}_1 ) {\bf b}_{+} } 
     + e^{ i ( {\bf q} - {\bf k}_2 ) {\bf b}_{+} } \right] 
\phi_{{\bf q}, {\bf k}_1 + {\bf k}_2 - {\bf q} } \, , 
\label{3UV2d:eq:atwenty} 
\end{equation}
where $\varepsilon_{\bf k}$ is the one-particle dispersion
\begin{equation}
\varepsilon_{\bf k} = - 2t \left( \cos{k_x} + \cos{k_y} \right) - 2 t_{\perp} \cos{k_z} \: ,  
\label{3UV3d:eq:atwentyone}
\end{equation}
and ${\bf b}_{+} = + {\bf x}$ or $+ {\bf y}$ are {\em two} nearest-neighbor vectors in the $xy$ plane. Note that ${\bf q}{\bf x} = q_x$ and so on. Total momentum ${\bf P} = {\bf k}_1 + {\bf k}_2$ is conserved, which allows writing the wave functions under ${\bf q}$ integrals as $\phi_{{\bf q}, {\bf P} - {\bf q} }$ and treating ${\bf P}$ as a parameter. Next, introduce three auxiliary functions
\begin{equation}
\Phi_{0}({\bf P}) = \frac{1}{N} \sum_{\bf q} \phi_{ {\bf q} , {\bf P} - {\bf q} } \: ,    
\hspace{0.5cm}
\Phi_{\bf x}({\bf P}) =  
\frac{1}{N} \sum_{\bf q} e^{ i q_x } \, \phi_{ {\bf q} , {\bf P} - {\bf q} } \: ,    
\hspace{0.5cm}
\Phi_{\bf y}({\bf P}) =  
\frac{1}{N} \sum_{\bf q} e^{ i q_y } \, \phi_{ {\bf q} , {\bf P} - {\bf q} } \: .    
\label{twopart:eq:atwentytwo}
\end{equation}
The wave function follows from Eq.~(\ref{3UV2d:eq:atwenty})
\begin{equation}
\phi_{{\bf k}_1 {\bf k}_2} = \frac{ U \Phi_{0}({\bf P}) 
- V ( e^{ - i k_{1x} } + e^{ - i k_{2x} } ) \Phi_{\bf x}({\bf P}) 
- V ( e^{ - i k_{1y} } + e^{ - i k_{2y} } ) \Phi_{\bf y}({\bf P}) }
     { E_2 - \varepsilon_{{\bf k}_1} - \varepsilon_{{\bf k}_2} } \: . 
\label{3UV2d:eq:atwentyfour} 
\end{equation}
Substituting this solution back into definitions (\ref{twopart:eq:atwentytwo}) one obtains a system of three linear equations for $\Phi$s
\begin{eqnarray}
\Phi_{0}     & = & \Phi_{0} \frac{U}{N} \sum_{\bf q} 
\frac{1}{ E_2 - \varepsilon_{\bf q} - \varepsilon_{{\bf P} - {\bf q}} } 
- \Phi_{\bf x} \frac{V}{N} \sum_{\bf q} 
\frac{ e^{ - i q_x } + e^{ - i ( P_x - q_x ) } }
{ E_2 - \varepsilon_{\bf q} - \varepsilon_{{\bf P} - {\bf q}} } 
- \Phi_{\bf y} \frac{V}{N} \sum_{\bf q} 
\frac{ e^{ - i q_y } + e^{ - i ( P_y - q_y ) } }
{ E_2 - \varepsilon_{\bf q} - \varepsilon_{{\bf P} - {\bf q}} }        \: ,
\label{3UV2d:eq:atwentyfive} \\
\Phi_{\bf x} & = & \Phi_{0} \frac{U}{N} \sum_{\bf q} 
\frac{ e^{ i q_x } }
{ E_2 - \varepsilon_{\bf q} - \varepsilon_{{\bf P} - {\bf q}} } 
- \Phi_{\bf x} \frac{V}{N} \sum_{\bf q} 
\frac{ e^{ i q_x } [ e^{ - i q_x } + e^{ - i ( P_x - q_x ) } ] }
{ E_2 - \varepsilon_{\bf q} - \varepsilon_{{\bf P} - {\bf q}} } 
- \Phi_{\bf y} \frac{V}{N} \sum_{\bf q} 
\frac{ e^{ i q_x } [ e^{ - i q_y } + e^{ - i ( P_y - q_y ) } ] }
{ E_2 - \varepsilon_{\bf q} - \varepsilon_{{\bf P} - {\bf q}} }        \: ,
\label{3UV2d:eq:atwentysix} \\
\Phi_{\bf y} & = & \Phi_{0} \frac{U}{N} \sum_{\bf q} 
\frac{ e^{ i q_y } }
{ E_2 - \varepsilon_{\bf q} - \varepsilon_{{\bf P} - {\bf q}} } 
- \Phi_{\bf x} \frac{V}{N} \sum_{\bf q} 
\frac{ e^{ i q_y } [ e^{ - i q_x } + e^{ - i ( P_x - q_x ) } ] }
{ E_2 - \varepsilon_{\bf q} - \varepsilon_{{\bf P} - {\bf q}} } 
- \Phi_{\bf y} \frac{V}{N} \sum_{\bf q} 
\frac{ e^{ i q_y } [ e^{ - i q_y } + e^{ - i ( P_y - q_y ) } ] }
{ E_2 - \varepsilon_{\bf q} - \varepsilon_{{\bf P} - {\bf q}} }        \: .
\label{3UV2d:eq:atwentyseven}
\end{eqnarray}
Shifting integration variables $q'_j = q_j - \frac{1}{2} P_j$ and changing the sign of the denominator, the last system can be rewritten as follows:
\begin{equation}
\left[ \begin{array}{ccc}
U M_{000} + 1  & - 2 V M_{100}                  & - 2 V M_{010}                 \\
U M_{100}      & - V ( M_{000} + M_{200} ) + 1  & - 2 V M_{110}                 \\
U M_{010}      & - 2 V M_{110}                  & - V ( M_{000} + M_{020} ) + 1  
\end{array} \right] 
\left[ \begin{array}{c}
              \Phi_{0}       \\
e^{-i(P_x/2)} \Phi_{\bf x}   \\
e^{-i(P_y/2)} \Phi_{\bf y}
\end{array} \right] = 0 \: ,
\label{3UV2d:eq:oneforty} 
\end{equation}
where 
\begin{equation}
M_{nml} \equiv \frac{1}{N} \sum_{\bf q} \frac{\cos{(nq_x)}\cos{(mq_y)}\cos{(lq_z)}}
{\vert E_2 \vert - a \cos{q_x} - b \cos{q_y} - c \cos{q_z} } \: ,  
\label{3UV2d:eq:onefortyone} 
\end{equation}
and $a \equiv 4 t \cos{\frac{P_x}{2}} \geq 0$, $b \equiv 4 t \cos{\frac{P_y}{2}} \geq 0$, and $c \equiv 4 t_{\perp} \cos{\frac{P_z}{2}} \geq 0$. If the particles are bound into a pair, the total energy $E_2 < - a - b - c$, and the integrals (\ref{3UV2d:eq:onefortyone}) are well defined. The consistency condition of Eq.~(\ref{3UV2d:eq:oneforty}) is equivalent to Eq.~(\ref{3UV3d:eq:three}) of the main text. 

Although the triple integrals (\ref{3UV2d:eq:onefortyone}) and can always be computed numerically, solving Eq.~(\ref{3UV2d:eq:oneforty}) for $E$ at an arbitrary ${\bf P}$ requires many evaluations of $M_{nml}$, which significantly slows down computation. It is much more efficient to perform two integrations analytically using the following formulae:
\begin{equation}
\int^{\pi}_{0} \!\!\! \int^{\pi}_{0} 
\! \frac{d\eta d\zeta}{\pi^2} \frac{1}
{ \alpha - \beta \cos{\eta} - \gamma \cos{\zeta} } 
= \frac{2}{\pi \sqrt{ \alpha^2 - (\beta - \gamma)^2 } } \: {\bf K}(\kappa) \: ,  
\label{3UV2d:eq:onefortytwo} 
\end{equation}
\begin{equation}
\int^{\pi}_{0} \!\!\! \int^{\pi}_{0} 
\! \frac{d\eta d\zeta}{\pi^2} \frac{\cos{\eta}\cos{\zeta}}
{ \alpha - \beta \cos{\eta} - \gamma \cos{\zeta} } 
= \frac{1}{\pi \kappa \sqrt{\beta\gamma} } 
\left\{ ( 2 - \kappa^2) \: {\bf K}(\kappa) - 2 {\bf E}(\kappa) \right\} \: .  
\label{3UV2d:eq:onefortyfive} 
\end{equation}
Here ${\bf K}(\kappa)$ and ${\bf E}(\kappa)$ are the complete elliptic integrals of the first and second types and their modulus is 
\begin{equation}
\kappa = \sqrt{\frac{4\beta\gamma}{\alpha^2 - (\beta - \gamma)^2}} \: .  
\label{3UV2d:eq:onefortyeight} 
\end{equation}
The third integration is performed numerically. For example, $M_{020}$ reduces to ($\eta = x$, $\zeta = z$) 
\begin{equation}
M_{020} = \int^{\pi}_{-\pi} \frac{dy}{2\pi} 
\frac{2 \cos{(2q_y)} {\bf K}(\kappa_y) } 
{ \pi \sqrt{ ( \vert E_2 \vert - b \cos{q_y} )^2 - ( a - c )^2 } }  \: ,  
\hspace{0.5cm}
\kappa_y = \sqrt{\frac{ 4 a c }
{ ( \vert E_2 \vert - b \cos{q_y} )^2 - ( a - c )^2 } } \: .  
\label{3UV2d:eq:atwentyeight} 
\end{equation}
The remaining integral is easy to compute numerically.

\subsection{\label{3UV3d:sec:app:b}
Three fermions with total spin $S = 1/2$.
}

Unlike the two-body problem, the three-body $UV$ problem is not exactly solvable. However, it is reducible to a system of integral equations of one variable, which can then be solved numerically. To preselect fermion states with a total spin $S = 1/2$, the wave function should be anti-symmetrized with respect to two arguments, for example ${\bf q}_1$ and ${\bf q}_2$. Such an anti-symmetrized Schr\"odinger equation has the form
\begin{eqnarray}
[ E - \varepsilon_{{\bf q}_1} - \varepsilon_{{\bf q}_2} - \varepsilon_{{\bf q}_3} ] 
\phi_{ {\bf q}_1 , {\bf q}_2 , {\bf q}_3 }
& = & \frac{U}{2N} \sum_{\bf k} \left\{ 
  \phi_{ {\bf q}_1 , {\bf k} , {\bf q}_2 + {\bf q}_3 - {\bf k} }
- \phi_{ {\bf q}_2 , {\bf k} , {\bf q}_1 + {\bf q}_3 - {\bf k} }  
+ \phi_{ {\bf q}_1 + {\bf q}_3 - {\bf k} , {\bf q}_2 , {\bf k} }
- \phi_{ {\bf q}_2 + {\bf q}_3 - {\bf k} , {\bf q}_1 , {\bf k} } 
\right\}
\nonumber \\
& & - \frac{V}{N} \sum_{\bf k} \left\{
\left[ \cos( q_{1x} - k_{x} ) + \cos( q_{1y} - k_{y} ) \right]
\phi_{ {\bf k} , {\bf q}_1 + {\bf q}_2 - {\bf k} , {\bf q}_3 } -
\right.
\nonumber \\
&& \hspace{1.5cm} 
\left[ \cos( q_{2x} - k_{x} ) + \cos( q_{2y} - k_{y} ) \right]
\phi_{ {\bf k} , {\bf q}_1 + {\bf q}_2 - {\bf k} , {\bf q}_3 } +
\nonumber \\ 
&& \hspace{1.5cm} 
\left[ \cos( q_{2x} - k_{x} ) + \cos( q_{2y} - k_{y} ) \right]
\phi_{ {\bf q}_1 , {\bf k} , {\bf q}_2 + {\bf q}_3 - {\bf k} } -
\nonumber \\
&& \hspace{1.5cm} 
\left[ \cos( q_{1x} - k_{x} ) + \cos( q_{1y} - k_{y} ) \right]
\phi_{ {\bf q}_2 , {\bf k} , {\bf q}_1 + {\bf q}_3 - {\bf k} } +
\nonumber \\
&& \hspace{1.5cm} 
\left[ \cos( q_{3x} - k_{x} ) + \cos( q_{3y} - k_{y} ) \right]
\phi_{ {\bf q}_1 + {\bf q}_3 - {\bf k} , {\bf q}_2 , {\bf k} } -
\nonumber \\
&& \hspace{1.45cm} \left.
\left[ \cos( q_{3x} - k_{x} ) + \cos( q_{3y} - k_{y} ) \right]
\phi_{ {\bf q}_2 + {\bf q}_3 - {\bf k} , {\bf q}_1 , {\bf k} }   
\right\}  \: .
\label{3UV2d:eq:atwentynine} 
\end{eqnarray}
By shifting the variable ${\bf k}$ and making use of the antisymmetry with respect to the first two arguments, the right-hand-side of Eq.~(\ref{3UV2d:eq:atwentynine}) can be transformed to a form where the first argument of $\phi$ is always ${\bf k}$: 
\begin{eqnarray}
& & [ E - \varepsilon_{{\bf q}_1} - \varepsilon_{{\bf q}_2} - \varepsilon_{{\bf q}_3} ] 
\phi_{ {\bf q}_1 , {\bf q}_2 , {\bf q}_3 }
= \frac{U}{N} \sum_{\bf k} \left\{ 
  \phi_{ {\bf k} , {\bf q}_2 , {\bf q}_1 + {\bf q}_3 - {\bf k} }
- \phi_{ {\bf k} , {\bf q}_1 , {\bf q}_2 + {\bf q}_3 - {\bf k} }  
\right\}
\nonumber \\
& & \hspace{1.5cm} 
- \frac{V}{N} \sum_{\bf k} \left\{
\left[ ( \cos{q_{1x}} - \cos{q_{2x}} ) \cos{k_x} + ( \cos{q_{1y}} - \cos{q_{2y}} ) \cos{k_y} + 
\right. \right. 
\nonumber \\
& & \hspace{3.0cm} 
\left. ( \sin{q_{1x}} - \sin{q_{2x}} ) \sin{k_x} + ( \sin{q_{1y}} - \sin{q_{2y}} ) \sin{k_y}
\right] 
\phi_{ {\bf k} , {\bf q}_1 + {\bf q}_2 - {\bf k} , {\bf q}_3 } +
\nonumber \\
& & \hspace{3.0cm} 
2 \left[ \cos{q_{1x}} \cos{k_{x}} + \cos{q_{1y}} \cos{k_{y}} + 
          \sin{q_{1x}} \sin{k_{x}} + \sin{q_{1y}} \sin{k_{y}} \right]
\phi_{ {\bf k} , {\bf q}_2 , {\bf q}_1 + {\bf q}_3 - {\bf k} } -
\nonumber \\ 
& & \hspace{3.0cm} \left. 
2 \left[ \cos{q_{2x}} \cos{k_{x}} + \cos{q_{2y}} \cos{k_{y}} + 
         \sin{q_{2x}} \sin{k_{x}} + \sin{q_{2y}} \sin{k_{y}} \right]
\phi_{ {\bf k} , {\bf q}_1 , {\bf q}_2 + {\bf q}_3 - {\bf k} }  \right\} \: . 
\label{3UV2d:eq:athirty} 
\end{eqnarray}
Next, usage is made of momentum conservation. Since total momentum ${\bf P} = {\bf q}_1 + {\bf q}_2 + {\bf q}_3$ is conserved, the wave function can be written as depending on ${\bf k}$ and only one ${\bf q}$. For example: $\phi_{ {\bf k} , {\bf q}_1 + {\bf q}_2 - {\bf k} , {\bf q}_3 } = \phi_{ {\bf k} , {\bf P} - {\bf q}_3 - {\bf k} , {\bf q}_3 }$, and so on. To illustrate the reduction procedure, consider the integral in the first term of Eq.~(\ref{3UV2d:eq:athirty}) 
\begin{equation}
\frac{1}{N} \sum_{\bf k} \left\{ 
  \phi_{ {\bf k} , {\bf q}_2 , {\bf q}_1 + {\bf q}_3 - {\bf k} }
- \phi_{ {\bf k} , {\bf q}_1 , {\bf q}_2 + {\bf q}_3 - {\bf k} }  
\right\} = 
\frac{1}{N} \sum_{\bf k} \left\{ 
  \phi_{ {\bf k} , {\bf q}_2 , {\bf P} - {\bf q}_2 - {\bf k} }
- \phi_{ {\bf k} , {\bf q}_1 , {\bf P} - {\bf q}_1 - {\bf k} }  
\right\} \equiv
F({\bf q}_2) - F({\bf q}_1) \: .
\label{3UV2d:eq:athirtyone} 
\end{equation}
Notice that it involves a new auxiliary function $F$ of only {\em one} variable but taken at two different arguments. Considering the right-hand-side of Eq.~(\ref{3UV2d:eq:athirty}), one defines nine auxiliary functions 
\begin{eqnarray}
F_1({\bf q}) & = & \frac{1}{N} \sum_{\bf k}        
   \phi_{{\bf k},{\bf q},{\bf P} - {\bf q} - {\bf k}} \: ,
\label{3UV2d:eq:onetwnetyone}    \\   
F_2({\bf q}) & = & \frac{1}{N} \sum_{\bf k} \cos{(k_x)} \,       
   \phi_{{\bf k},{\bf P} - {\bf q} - {\bf k},{\bf q}} \: ,
\label{3UV2d:eq:onetwentytwo}  \\
F_3({\bf q}) & = & \frac{1}{N} \sum_{\bf k} \cos{(k_y)} \,       
   \phi_{{\bf k},{\bf P} - {\bf q} - {\bf k},{\bf q}} \: ,
\label{3UV2d:eq:onetwentythree}   \\
F_4({\bf q}) & = & \frac{1}{N} \sum_{\bf k} \cos{(k_x)} \,       
   \phi_{{\bf k},{\bf q},{\bf P} - {\bf q} - {\bf k}} \: ,
\label{3UV2d:eq:onetwentyfour}    \\   
F_5({\bf q}) & = & \frac{1}{N} \sum_{\bf k} \cos{(k_y)} \,       
   \phi_{{\bf k},{\bf q},{\bf P} - {\bf q} - {\bf k}} \: ,
\label{3UV2d:eq:onetwentyfive}  \\
F_6({\bf q}) & = & \frac{1}{N} \sum_{\bf k} \sin{(k_x)} \,       
   \phi_{{\bf k},{\bf P} - {\bf q} - {\bf k},{\bf q}} \: ,
\label{3UV2d:eq:onetwentysix}  \\
F_7({\bf q}) & = & \frac{1}{N} \sum_{\bf k} \sin{(k_y)} \,       
   \phi_{{\bf k},{\bf P} - {\bf q} - {\bf k},{\bf q}} \: ,
\label{3UV2d:eq:onetwentyseven}   \\
F_8({\bf q}) & = & \frac{1}{N} \sum_{\bf k} \sin{(k_x)} \,       
   \phi_{{\bf k},{\bf q},{\bf P} - {\bf q} - {\bf k}} \: ,
\label{3UV2d:eq:onetwentyeight}   \\
F_9({\bf q}) & = & \frac{1}{N} \sum_{\bf k} \sin{(k_y)} \,       
   \phi_{{\bf k},{\bf q},{\bf P} - {\bf q} - {\bf k}} \: .
\label{3UV2d:eq:onetwentynine}
\end{eqnarray}
The Schr\"odinger equation, Eq.~(\ref{3UV2d:eq:athirty}), becomes:
\begin{eqnarray}
& & [ E - \varepsilon_{{\bf q}_1} - \varepsilon_{{\bf q}_2} - \varepsilon_{{\bf q}_3} ] 
\phi_{ {\bf q}_1 , {\bf q}_2 , {\bf q}_3 } 
= U \left[ F_1({\bf q}_2) - F_1({\bf q}_1) \right]
\nonumber \\
& & \hspace{2.3cm} 
- V \left[ ( \cos{q_{1x}} - \cos{q_{2x}} ) F_2({\bf q}_3) + 
           ( \cos{q_{1y}} - \cos{q_{2y}} ) F_3({\bf q}_3) + 
\right. 
\nonumber \\
& & \hspace{3.0cm} 
       ( \sin{q_{1x}} - \sin{q_{2x}} ) F_6({\bf q}_3) + 
       ( \sin{q_{1y}} - \sin{q_{2y}} ) F_7({\bf q}_3) +
\nonumber \\
& & \hspace{3.0cm} 
2 \cos{q_{1x}} F_4({\bf q}_2) + 2 \cos{q_{1y}} F_5({\bf q}_2) + 
2 \sin{q_{1x}} F_8({\bf q}_2) + 2 \sin{q_{1y}} F_9({\bf q}_2) - 
\nonumber \\ 
& & \hspace{3.0cm} \left. 
2 \cos{q_{2x}} F_4({\bf q}_1) + 2 \cos{q_{2y}} F_5({\bf q}_1) + 
2 \sin{q_{2x}} F_8({\bf q}_1) + 2 \sin{q_{2y}} F_9({\bf q}_1) 
\right] \: . 
\label{3UV2d:eq:athirtytwo} 
\end{eqnarray}
Now wave function $\phi_{ {\bf q}_1 , {\bf q}_2 , {\bf q}_3 }$ can be expressed as a linear combination of $F$, and substituted back into the definitions (\ref{3UV2d:eq:onetwnetyone})-(\ref{3UV2d:eq:onetwentynine}). Taking into account that ${\bf q}_3 = {\bf P} - {\bf q}_1 - {\bf q}_2$, one derives a system of nine coupled integral equations given below in Eqs.~(\ref{3UV2d:eq:onethirtyone})-(\ref{3UV2d:eq:onethirtynine}). They can be solved numerically by discretizing the Brillouin zone and converting into a matrix equation. Then the system's energy $E$ can be found via eigenvalue search. $E$ is adjusted until the matrix equation acquires an eigenvalue equal to 1. 
\begin{eqnarray}
F_1({\bf q}) & = & \frac{U}{N} \sum_{\bf k} 
\frac{F_1({\bf q}) - F_1({\bf k})}
{E - \varepsilon({\bf k}) - \varepsilon({\bf q}) - \varepsilon({\bf P}-{\bf q}-{\bf k})}
\nonumber \\
       &   & - \frac{V}{N} \sum_{\bf k}
\frac{[\cos(P_x - q_x - k_x) - \cos(q_x)] \, F_2({\bf k}) }
{E - \varepsilon({\bf k}) - \varepsilon({\bf q}) - \varepsilon({\bf P}-{\bf q}-{\bf k})}    
\nonumber \\
       &   & - \frac{V}{N} \sum_{\bf k}
\frac{[\cos(P_y - q_y - k_y) - \cos(q_y)] \, F_3({\bf k}) }
{E - \varepsilon({\bf k}) - \varepsilon({\bf q}) - \varepsilon({\bf P}-{\bf q}-{\bf k})}    
\nonumber \\
       &   & - \frac{V}{N} \sum_{\bf k}
\frac{2 \cos(k_x) \, F_4({\bf q}) - 2 \cos(q_x) \, F_4({\bf k}) }
{E - \varepsilon({\bf k}) - \varepsilon({\bf q}) - \varepsilon({\bf P}-{\bf q}-{\bf k})}    
\nonumber \\
       &   & - \frac{V}{N} \sum_{\bf k}
\frac{2 \cos(k_y) \, F_5({\bf q}) - 2 \cos(q_y) \, F_5({\bf k}) }
{E - \varepsilon({\bf k}) - \varepsilon({\bf q}) - \varepsilon({\bf P}-{\bf q}-{\bf k})}    
\nonumber \\
       &   & - \frac{V}{N} \sum_{\bf k}
\frac{[\sin(P_x - q_x - k_x) - \sin(q_x)] \, F_6({\bf k}) }
{E - \varepsilon({\bf k}) - \varepsilon({\bf q}) - \varepsilon({\bf P}-{\bf q}-{\bf k})}    
\nonumber \\
       &   & - \frac{V}{N} \sum_{\bf k}
\frac{[\sin(P_y - q_y - k_y) - \sin(q_y)] \, F_7({\bf k}) }
{E - \varepsilon({\bf k}) - \varepsilon({\bf q}) - \varepsilon({\bf P}-{\bf q}-{\bf k})}    
\nonumber \\
       &   & - \frac{V}{N} \sum_{\bf k}
\frac{2 \sin(k_x) \, F_8({\bf q}) - 2 \sin(q_x) \, F_8({\bf k}) }
{E - \varepsilon({\bf k}) - \varepsilon({\bf q}) - \varepsilon({\bf P}-{\bf q}-{\bf k})}    
\nonumber \\
       &   & - \frac{V}{N} \sum_{\bf k}
\frac{2 \sin(k_y) \, F_9({\bf q}) - 2 \sin(q_y) \, F_9({\bf k}) }
{E - \varepsilon({\bf k}) - \varepsilon({\bf q}) - \varepsilon({\bf P}-{\bf q}-{\bf k})}  \: ,  
\label{3UV2d:eq:onethirtyone} 
\end{eqnarray}
\begin{eqnarray}
F_2({\bf q}) & = & \frac{U}{N} \sum_{\bf k} 
\frac{[\cos(P_x - q_x - k_x) - \cos(k_x)] \, F_1({\bf k})}
{E - \varepsilon({\bf k}) - \varepsilon({\bf q}) - \varepsilon({\bf P}-{\bf q}-{\bf k})}
\nonumber \\
       &   & - \frac{V}{N} \sum_{\bf k}
\frac{\cos(k_x) [\cos(k_x) - \cos(P_x - q_x - k_x)] \, F_2({\bf q}) }
{E - \varepsilon({\bf k}) - \varepsilon({\bf q}) - \varepsilon({\bf P}-{\bf q}-{\bf k})}    
\nonumber \\
       &   & - \frac{V}{N} \sum_{\bf k}
\frac{\cos(k_x) [\cos(k_y) - \cos(P_y - q_y - k_y)] \, F_3({\bf q}) }
{E - \varepsilon({\bf k}) - \varepsilon({\bf q}) - \varepsilon({\bf P}-{\bf q}-{\bf k})}    
\nonumber \\
       &   & - \frac{V}{N} \sum_{\bf k}
\frac{2 \cos(P_x - q_x - k_x) [\cos(P_x - q_x - k_x) - \cos(k_x)] \, F_4({\bf k}) }
{E - \varepsilon({\bf k}) - \varepsilon({\bf q}) - \varepsilon({\bf P}-{\bf q}-{\bf k})}    
\nonumber \\
       &   & - \frac{V}{N} \sum_{\bf k}
\frac{2 \cos(P_y - q_y - k_y) [\cos(P_x - q_x - k_x) - \cos(k_x)] \, F_5({\bf k}) }
{E - \varepsilon({\bf k}) - \varepsilon({\bf q}) - \varepsilon({\bf P}-{\bf q}-{\bf k})}    
\nonumber \\
       &   & - \frac{V}{N} \sum_{\bf k}
\frac{\cos(k_x) [\sin(k_x) - \sin(P_x - q_x - k_x)] \, F_6({\bf q}) }
{E - \varepsilon({\bf k}) - \varepsilon({\bf q}) - \varepsilon({\bf P}-{\bf q}-{\bf k})}    
\nonumber \\
       &   & - \frac{V}{N} \sum_{\bf k}
\frac{\cos(k_x) [\sin(k_y) - \sin(P_y - q_y - k_y)] \, F_7({\bf q}) }
{E - \varepsilon({\bf k}) - \varepsilon({\bf q}) - \varepsilon({\bf P}-{\bf q}-{\bf k})}    
\nonumber \\
       &   & - \frac{V}{N} \sum_{\bf k}
\frac{2 \sin(P_x - q_x - k_x) [\cos(P_x - q_x - k_x) - \cos(k_x)] \, F_8({\bf k}) }
{E - \varepsilon({\bf k}) - \varepsilon({\bf q}) - \varepsilon({\bf P}-{\bf q}-{\bf k})}    
\nonumber \\
       &   & - \frac{V}{N} \sum_{\bf k}
\frac{2 \sin(P_y - q_y - k_y) [\cos(P_x - q_x - k_x) - \cos(k_x)] \, F_9({\bf k}) }
{E - \varepsilon({\bf k}) - \varepsilon({\bf q}) - \varepsilon({\bf P}-{\bf q}-{\bf k})}  \: ,  
\label{3UV2d:eq:onethirtytwo} 
\end{eqnarray}
\begin{eqnarray}
F_3({\bf q}) & = & \frac{U}{N} \sum_{\bf k} 
\frac{[\cos(P_y - q_y - k_y) - \cos(k_y)] \, F_1({\bf k})}
{E - \varepsilon({\bf k}) - \varepsilon({\bf q}) - \varepsilon({\bf P}-{\bf q}-{\bf k})}
\nonumber \\
       &   & - \frac{V}{N} \sum_{\bf k}
\frac{\cos(k_y) [\cos(k_x) - \cos(P_x - q_x - k_x)] \, F_2({\bf q}) }
{E - \varepsilon({\bf k}) - \varepsilon({\bf q}) - \varepsilon({\bf P}-{\bf q}-{\bf k})}    
\nonumber \\
       &   & - \frac{V}{N} \sum_{\bf k}
\frac{\cos(k_y) [\cos(k_y) - \cos(P_y - q_y - k_y)] \, F_3({\bf q}) }
{E - \varepsilon({\bf k}) - \varepsilon({\bf q}) - \varepsilon({\bf P}-{\bf q}-{\bf k})}    
\nonumber \\
       &   & - \frac{V}{N} \sum_{\bf k}
\frac{2 \cos(P_x - q_x - k_x) [\cos(P_y - q_y - k_y) - \cos(k_y)] \, F_4({\bf k}) }
{E - \varepsilon({\bf k}) - \varepsilon({\bf q}) - \varepsilon({\bf P}-{\bf q}-{\bf k})}    
\nonumber \\
       &   & - \frac{V}{N} \sum_{\bf k}
\frac{2 \cos(P_y - q_y - k_y) [\cos(P_y - q_y - k_y) - \cos(k_y)] \, F_5({\bf k}) }
{E - \varepsilon({\bf k}) - \varepsilon({\bf q}) - \varepsilon({\bf P}-{\bf q}-{\bf k})}    
\nonumber \\
       &   & - \frac{V}{N} \sum_{\bf k}
\frac{\cos(k_y) [\sin(k_x) - \sin(P_x - q_x - k_x)] \, F_6({\bf q}) }
{E - \varepsilon({\bf k}) - \varepsilon({\bf q}) - \varepsilon({\bf P}-{\bf q}-{\bf k})}    
\nonumber \\
       &   & - \frac{V}{N} \sum_{\bf k}
\frac{\cos(k_y) [\sin(k_y) - \sin(P_y - q_y - k_y)] \, F_7({\bf q}) }
{E - \varepsilon({\bf k}) - \varepsilon({\bf q}) - \varepsilon({\bf P}-{\bf q}-{\bf k})}    
\nonumber \\
       &   & - \frac{V}{N} \sum_{\bf k}
\frac{2 \sin(P_x - q_x - k_x) [\cos(P_y - q_y - k_y) - \cos(k_y)] \, F_8({\bf k}) }
{E - \varepsilon({\bf k}) - \varepsilon({\bf q}) - \varepsilon({\bf P}-{\bf q}-{\bf k})}    
\nonumber \\
       &   & - \frac{V}{N} \sum_{\bf k}
\frac{2 \sin(P_y - q_y - k_y) [\cos(P_y - q_y - k_y) - \cos(k_y)] \, F_9({\bf k}) }
{E - \varepsilon({\bf k}) - \varepsilon({\bf q}) - \varepsilon({\bf P}-{\bf q}-{\bf k})}  \: ,  
\label{3UV2d:eq:onethirtythree} 
\end{eqnarray}
\begin{eqnarray}
F_4({\bf q}) & = & \frac{U}{N} \sum_{\bf k} 
\frac{\cos(k_x) \, F_1({\bf q}) - \cos(k_x) \, F_1({\bf k})}
{E - \varepsilon({\bf k}) - \varepsilon({\bf q}) - \varepsilon({\bf P}-{\bf q}-{\bf k})}
\nonumber \\
       &   & - \frac{V}{N} \sum_{\bf k}
\frac{\cos(P_x - q_x - k_x) [\cos(P_x - q_x - k_x) - \cos(q_x)] \, F_2({\bf k}) }
{E - \varepsilon({\bf k}) - \varepsilon({\bf q}) - \varepsilon({\bf P}-{\bf q}-{\bf k})}    
\nonumber \\
       &   & - \frac{V}{N} \sum_{\bf k}
\frac{\cos(P_x - q_x - k_x) [\cos(P_y - q_y - k_y) - \cos(q_y)] \, F_3({\bf k}) }
{E - \varepsilon({\bf k}) - \varepsilon({\bf q}) - \varepsilon({\bf P}-{\bf q}-{\bf k})}    
\nonumber \\
       &   & - \frac{V}{N} \sum_{\bf k}
\frac{2 \cos^2(k_x) \, F_4({\bf q}) - 2 \cos(k_x) \cos(q_x) \, F_4({\bf k}) }
{E - \varepsilon({\bf k}) - \varepsilon({\bf q}) - \varepsilon({\bf P}-{\bf q}-{\bf k})}    
\nonumber \\
       &   & - \frac{V}{N} \sum_{\bf k}
\frac{2 \cos(k_x) \cos(k_y) \, F_5({\bf q}) - 2 \cos(k_x) \cos(q_y) \, F_5({\bf k}) }
{E - \varepsilon({\bf k}) - \varepsilon({\bf q}) - \varepsilon({\bf P}-{\bf q}-{\bf k})}    
\nonumber \\
       &   & - \frac{V}{N} \sum_{\bf k}
\frac{\cos(P_x - q_x - k_x) [\sin(P_x - q_x - k_x) - \sin(q_x)] \, F_6({\bf k}) }
{E - \varepsilon({\bf k}) - \varepsilon({\bf q}) - \varepsilon({\bf P}-{\bf q}-{\bf k})}    
\nonumber \\
       &   & - \frac{V}{N} \sum_{\bf k}
\frac{\cos(P_x - q_x - k_x) [\sin(P_y - q_y - k_y) - \sin(q_y)] \, F_7({\bf k}) }
{E - \varepsilon({\bf k}) - \varepsilon({\bf q}) - \varepsilon({\bf P}-{\bf q}-{\bf k})}    
\nonumber \\
       &   & - \frac{V}{N} \sum_{\bf k}
\frac{2 \cos(k_x) \sin(k_x) \, F_8({\bf q}) - 2 \cos(k_x) \sin(q_x) \, F_8({\bf k}) }
{E - \varepsilon({\bf k}) - \varepsilon({\bf q}) - \varepsilon({\bf P}-{\bf q}-{\bf k})}    
\nonumber \\
       &   & - \frac{V}{N} \sum_{\bf k}
\frac{2 \cos(k_x) \sin(k_y) \, F_9({\bf q}) - 2 \cos(k_x) \sin(q_y) \, F_9({\bf k}) }
{E - \varepsilon({\bf k}) - \varepsilon({\bf q}) - \varepsilon({\bf P}-{\bf q}-{\bf k})}  \: ,  
\label{3UV2d:eq:onethirtyfour} 
\end{eqnarray}
\begin{eqnarray}
F_5({\bf q}) & = & \frac{U}{N} \sum_{\bf k} 
\frac{\cos(k_y) \, F_1({\bf q}) - \cos(k_y) \, F_1({\bf k})}
{E - \varepsilon({\bf k}) - \varepsilon({\bf q}) - \varepsilon({\bf P}-{\bf q}-{\bf k})}
\nonumber \\
       &   & - \frac{V}{N} \sum_{\bf k}
\frac{\cos(P_y - q_y - k_y) [\cos(P_x - q_x - k_x) - \cos(q_x)] \, F_2({\bf k}) }
{E - \varepsilon({\bf k}) - \varepsilon({\bf q}) - \varepsilon({\bf P}-{\bf q}-{\bf k})}    
\nonumber \\
       &   & - \frac{V}{N} \sum_{\bf k}
\frac{\cos(P_y - q_y - k_y) [\cos(P_y - q_y - k_y) - \cos(q_y)] \, F_3({\bf k}) }
{E - \varepsilon({\bf k}) - \varepsilon({\bf q}) - \varepsilon({\bf P}-{\bf q}-{\bf k})}    
\nonumber \\
       &   & - \frac{V}{N} \sum_{\bf k}
\frac{2 \cos(k_y) \cos(k_x) \, F_4({\bf q}) - 2 \cos(k_y) \cos(q_x) \, F_4({\bf k}) }
{E - \varepsilon({\bf k}) - \varepsilon({\bf q}) - \varepsilon({\bf P}-{\bf q}-{\bf k})}    
\nonumber \\
       &   & - \frac{V}{N} \sum_{\bf k}
\frac{2 \cos^2(k_y) \, F_5({\bf q}) - 2 \cos(k_y) \cos(q_y) \, F_5({\bf k}) }
{E - \varepsilon({\bf k}) - \varepsilon({\bf q}) - \varepsilon({\bf P}-{\bf q}-{\bf k})}    
\nonumber \\
       &   & - \frac{V}{N} \sum_{\bf k}
\frac{\cos(P_y - q_y - k_y) [\sin(P_x - q_x - k_x) - \sin(q_x)] \, F_6({\bf k}) }
{E - \varepsilon({\bf k}) - \varepsilon({\bf q}) - \varepsilon({\bf P}-{\bf q}-{\bf k})}    
\nonumber \\
       &   & - \frac{V}{N} \sum_{\bf k}
\frac{\cos(P_y - q_y - k_y) [\sin(P_y - q_y - k_y) - \sin(q_y)] \, F_7({\bf k}) }
{E - \varepsilon({\bf k}) - \varepsilon({\bf q}) - \varepsilon({\bf P}-{\bf q}-{\bf k})}    
\nonumber \\
       &   & - \frac{V}{N} \sum_{\bf k}
\frac{2 \cos(k_y) \sin(k_x) \, F_8({\bf q}) - 2 \cos(k_y) \sin(q_x) \, F_8({\bf k}) }
{E - \varepsilon({\bf k}) - \varepsilon({\bf q}) - \varepsilon({\bf P}-{\bf q}-{\bf k})}    
\nonumber \\
       &   & - \frac{V}{N} \sum_{\bf k}
\frac{2 \cos(k_y) \sin(k_y) \, F_9({\bf q}) - 2 \cos(k_y) \sin(q_y) \, F_9({\bf k}) }
{E - \varepsilon({\bf k}) - \varepsilon({\bf q}) - \varepsilon({\bf P}-{\bf q}-{\bf k})}  \: ,  
\label{3UV2d:eq:onethirtyfive} 
\end{eqnarray}
\begin{eqnarray}
F_6({\bf q}) & = & \frac{U}{N} \sum_{\bf k} 
\frac{[\sin(P_x - q_x - k_x) - \sin(k_x)] \, F_1({\bf k})}
{E - \varepsilon({\bf k}) - \varepsilon({\bf q}) - \varepsilon({\bf P}-{\bf q}-{\bf k})}
\nonumber \\
       &   & - \frac{V}{N} \sum_{\bf k}
\frac{\sin(k_x) [\cos(k_x) - \cos(P_x - q_x - k_x)] \, F_2({\bf q}) }
{E - \varepsilon({\bf k}) - \varepsilon({\bf q}) - \varepsilon({\bf P}-{\bf q}-{\bf k})}    
\nonumber \\
       &   & - \frac{V}{N} \sum_{\bf k}
\frac{\sin(k_x) [\cos(k_y) - \cos(P_y - q_y - k_y)] \, F_3({\bf q}) }
{E - \varepsilon({\bf k}) - \varepsilon({\bf q}) - \varepsilon({\bf P}-{\bf q}-{\bf k})}    
\nonumber \\
       &   & - \frac{V}{N} \sum_{\bf k}
\frac{2 \cos(P_x - q_x - k_x) [\sin(P_x - q_x - k_x) - \sin(k_x)] \, F_4({\bf k}) }
{E - \varepsilon({\bf k}) - \varepsilon({\bf q}) - \varepsilon({\bf P}-{\bf q}-{\bf k})}    
\nonumber \\
       &   & - \frac{V}{N} \sum_{\bf k}
\frac{2 \cos(P_y - q_y - k_y) [\sin(P_x - q_x - k_x) - \sin(k_x)] \, F_5({\bf k}) }
{E - \varepsilon({\bf k}) - \varepsilon({\bf q}) - \varepsilon({\bf P}-{\bf q}-{\bf k})}    
\nonumber \\
       &   & - \frac{V}{N} \sum_{\bf k}
\frac{\sin(k_x) [\sin(k_x) - \sin(P_x - q_x - k_x)] \, F_6({\bf q}) }
{E - \varepsilon({\bf k}) - \varepsilon({\bf q}) - \varepsilon({\bf P}-{\bf q}-{\bf k})}    
\nonumber \\
       &   & - \frac{V}{N} \sum_{\bf k}
\frac{\sin(k_x) [\sin(k_y) - \sin(P_y - q_y - k_y)] \, F_7({\bf q}) }
{E - \varepsilon({\bf k}) - \varepsilon({\bf q}) - \varepsilon({\bf P}-{\bf q}-{\bf k})}    
\nonumber \\
       &   & - \frac{V}{N} \sum_{\bf k}
\frac{2 \sin(P_x - q_x - k_x) [\sin(P_x - q_x - k_x) - \sin(k_x)] \, F_8({\bf k}) }
{E - \varepsilon({\bf k}) - \varepsilon({\bf q}) - \varepsilon({\bf P}-{\bf q}-{\bf k})}    
\nonumber \\
       &   & - \frac{V}{N} \sum_{\bf k}
\frac{2 \sin(P_y - q_y - k_y) [\sin(P_x - q_x - k_x) - \sin(k_x)] \, F_9({\bf k}) }
{E - \varepsilon({\bf k}) - \varepsilon({\bf q}) - \varepsilon({\bf P}-{\bf q}-{\bf k})}  \: ,  
\label{3UV2d:eq:onethirtysix} 
\end{eqnarray}
\begin{eqnarray}
F_7({\bf q}) & = & \frac{U}{N} \sum_{\bf k} 
\frac{[\sin(P_y - q_y - k_y) - \sin(k_y)] \, F_1({\bf k})}
{E - \varepsilon({\bf k}) - \varepsilon({\bf q}) - \varepsilon({\bf P}-{\bf q}-{\bf k})}
\nonumber \\
       &   & - \frac{V}{N} \sum_{\bf k}
\frac{\sin(k_y) [\cos(k_x) - \cos(P_x - q_x - k_x)] \, F_2({\bf q}) }
{E - \varepsilon({\bf k}) - \varepsilon({\bf q}) - \varepsilon({\bf P}-{\bf q}-{\bf k})}    
\nonumber \\
       &   & - \frac{V}{N} \sum_{\bf k}
\frac{\sin(k_y) [\cos(k_y) - \cos(P_y - q_y - k_y)] \, F_3({\bf q}) }
{E - \varepsilon({\bf k}) - \varepsilon({\bf q}) - \varepsilon({\bf P}-{\bf q}-{\bf k})}    
\nonumber \\
       &   & - \frac{V}{N} \sum_{\bf k}
\frac{2 \cos(P_x - q_x - k_x) [\sin(P_y - q_y - k_y) - \sin(k_y)] \, F_4({\bf k}) }
{E - \varepsilon({\bf k}) - \varepsilon({\bf q}) - \varepsilon({\bf P}-{\bf q}-{\bf k})}    
\nonumber \\
       &   & - \frac{V}{N} \sum_{\bf k}
\frac{2 \cos(P_y - q_y - k_y) [\sin(P_y - q_y - k_y) - \sin(k_y)] \, F_5({\bf k}) }
{E - \varepsilon({\bf k}) - \varepsilon({\bf q}) - \varepsilon({\bf P}-{\bf q}-{\bf k})}    
\nonumber \\
       &   & - \frac{V}{N} \sum_{\bf k}
\frac{\sin(k_y) [\sin(k_x) - \sin(P_x - q_x - k_x)] \, F_6({\bf q}) }
{E - \varepsilon({\bf k}) - \varepsilon({\bf q}) - \varepsilon({\bf P}-{\bf q}-{\bf k})}    
\nonumber \\
       &   & - \frac{V}{N} \sum_{\bf k}
\frac{\sin(k_y) [\sin(k_y) - \sin(P_y - q_y - k_y)] \, F_7({\bf q}) }
{E - \varepsilon({\bf k}) - \varepsilon({\bf q}) - \varepsilon({\bf P}-{\bf q}-{\bf k})}    
\nonumber \\
       &   & - \frac{V}{N} \sum_{\bf k}
\frac{2 \sin(P_x - q_x - k_x) [\sin(P_y - q_y - k_y) - \sin(k_y)] \, F_8({\bf k}) }
{E - \varepsilon({\bf k}) - \varepsilon({\bf q}) - \varepsilon({\bf P}-{\bf q}-{\bf k})}    
\nonumber \\
       &   & - \frac{V}{N} \sum_{\bf k}
\frac{2 \sin(P_y - q_y - k_y) [\sin(P_y - q_y - k_y) - \sin(k_y)] \, F_9({\bf k}) }
{E - \varepsilon({\bf k}) - \varepsilon({\bf q}) - \varepsilon({\bf P}-{\bf q}-{\bf k})}  \: ,  
\label{3UV2d:eq:onethirtyseven} 
\end{eqnarray}
\begin{eqnarray}
F_8({\bf q}) & = & \frac{U}{N} \sum_{\bf k} 
\frac{\sin(k_x) \, F_1({\bf q}) - \sin(k_x) \, F_1({\bf k})}
{E - \varepsilon({\bf k}) - \varepsilon({\bf q}) - \varepsilon({\bf P}-{\bf q}-{\bf k})}
\nonumber \\
       &   & - \frac{V}{N} \sum_{\bf k}
\frac{\sin(P_x - q_x - k_x) [\cos(P_x - q_x - k_x) - \cos(q_x)] \, F_2({\bf k}) }
{E - \varepsilon({\bf k}) - \varepsilon({\bf q}) - \varepsilon({\bf P}-{\bf q}-{\bf k})}    
\nonumber \\
       &   & - \frac{V}{N} \sum_{\bf k}
\frac{\sin(P_x - q_x - k_x) [\cos(P_y - q_y - k_y) - \cos(q_y)] \, F_3({\bf k}) }
{E - \varepsilon({\bf k}) - \varepsilon({\bf q}) - \varepsilon({\bf P}-{\bf q}-{\bf k})}    
\nonumber \\
       &   & - \frac{V}{N} \sum_{\bf k}
\frac{2 \sin(k_x) \cos(k_x) \, F_4({\bf q}) - 2 \sin(k_x) \cos(q_x) \, F_4({\bf k}) }
{E - \varepsilon({\bf k}) - \varepsilon({\bf q}) - \varepsilon({\bf P}-{\bf q}-{\bf k})}    
\nonumber \\
       &   & - \frac{V}{N} \sum_{\bf k}
\frac{2 \sin(k_x) \cos(k_y) \, F_5({\bf q}) - 2 \sin(k_x) \cos(q_y) \, F_5({\bf k}) }
{E - \varepsilon({\bf k}) - \varepsilon({\bf q}) - \varepsilon({\bf P}-{\bf q}-{\bf k})}    
\nonumber \\
       &   & - \frac{V}{N} \sum_{\bf k}
\frac{\sin(P_x - q_x - k_x) [\sin(P_x - q_x - k_x) - \sin(q_x)] \, F_6({\bf k}) }
{E - \varepsilon({\bf k}) - \varepsilon({\bf q}) - \varepsilon({\bf P}-{\bf q}-{\bf k})}    
\nonumber \\
       &   & - \frac{V}{N} \sum_{\bf k}
\frac{\sin(P_x - q_x - k_x) [\sin(P_y - q_y - k_y) - \sin(q_y)] \, F_7({\bf k}) }
{E - \varepsilon({\bf k}) - \varepsilon({\bf q}) - \varepsilon({\bf P}-{\bf q}-{\bf k})}    
\nonumber \\
       &   & - \frac{V}{N} \sum_{\bf k}
\frac{2 \sin^2(k_x) \, F_8({\bf q}) - 2 \sin(k_x) \sin(q_x) \, F_8({\bf k}) }
{E - \varepsilon({\bf k}) - \varepsilon({\bf q}) - \varepsilon({\bf P}-{\bf q}-{\bf k})}    
\nonumber \\
       &   & - \frac{V}{N} \sum_{\bf k}
\frac{2 \sin(k_x) \sin(k_y) \, F_9({\bf q}) - 2 \sin(k_x) \sin(q_y) \, F_9({\bf k}) }
{E - \varepsilon({\bf k}) - \varepsilon({\bf q}) - \varepsilon({\bf P}-{\bf q}-{\bf k})}  \: ,  
\label{3UV2d:eq:onethirtyeight} 
\end{eqnarray}
\begin{eqnarray}
F_9({\bf q}) & = & \frac{U}{N} \sum_{\bf k} 
\frac{\sin(k_y) \, F_1({\bf q}) - \sin(k_y) \, F_1({\bf k})}
{E - \varepsilon({\bf k}) - \varepsilon({\bf q}) - \varepsilon({\bf P}-{\bf q}-{\bf k})}
\nonumber \\
       &   & - \frac{V}{N} \sum_{\bf k}
\frac{\sin(P_y - q_y - k_y) [\cos(P_x - q_x - k_x) - \cos(q_x)] \, F_2({\bf k}) }
{E - \varepsilon({\bf k}) - \varepsilon({\bf q}) - \varepsilon({\bf P}-{\bf q}-{\bf k})}    
\nonumber \\
       &   & - \frac{V}{N} \sum_{\bf k}
\frac{\sin(P_y - q_y - k_y) [\cos(P_y - q_y - k_y) - \cos(q_y)] \, F_3({\bf k}) }
{E - \varepsilon({\bf k}) - \varepsilon({\bf q}) - \varepsilon({\bf P}-{\bf q}-{\bf k})}    
\nonumber \\
       &   & - \frac{V}{N} \sum_{\bf k}
\frac{2 \sin(k_y) \cos(k_x) \, F_4({\bf q}) - 2 \sin(k_y) \cos(q_x) \, F_4({\bf k}) }
{E - \varepsilon({\bf k}) - \varepsilon({\bf q}) - \varepsilon({\bf P}-{\bf q}-{\bf k})}    
\nonumber \\
       &   & - \frac{V}{N} \sum_{\bf k}
\frac{2 \sin(k_y) \cos(k_y) \, F_5({\bf q}) - 2 \sin(k_y) \cos(q_y) \, F_5({\bf k}) }
{E - \varepsilon({\bf k}) - \varepsilon({\bf q}) - \varepsilon({\bf P}-{\bf q}-{\bf k})}    
\nonumber \\
       &   & - \frac{V}{N} \sum_{\bf k}
\frac{\sin(P_y - q_y - k_y) [\sin(P_x - q_x - k_x) - \sin(q_x)] \, F_6({\bf k}) }
{E - \varepsilon({\bf k}) - \varepsilon({\bf q}) - \varepsilon({\bf P}-{\bf q}-{\bf k})}    
\nonumber \\
       &   & - \frac{V}{N} \sum_{\bf k}
\frac{\sin(P_y - q_y - k_y) [\sin(P_y - q_y - k_y) - \sin(q_y)] \, F_7({\bf k}) }
{E - \varepsilon({\bf k}) - \varepsilon({\bf q}) - \varepsilon({\bf P}-{\bf q}-{\bf k})}    
\nonumber \\
       &   & - \frac{V}{N} \sum_{\bf k}
\frac{2 \sin(k_y) \sin(k_x) \, F_8({\bf q}) - 2 \sin(k_y) \sin(q_x) \, F_8({\bf k}) }
{E - \varepsilon({\bf k}) - \varepsilon({\bf q}) - \varepsilon({\bf P}-{\bf q}-{\bf k})}    
\nonumber \\
       &   & - \frac{V}{N} \sum_{\bf k}
\frac{2 \sin^2(k_y) \, F_9({\bf q}) - 2 \sin(k_y) \sin(q_y) \, F_9({\bf k}) }
{E - \varepsilon({\bf k}) - \varepsilon({\bf q}) - \varepsilon({\bf P}-{\bf q}-{\bf k})}  \: .  
\label{3UV2d:eq:onethirtynine} 
\end{eqnarray}

\end{widetext}

\end{document}